# Strain-induced metallization and defect suppression at zipper-like interdigitated atomically thin interfaces enabling high-efficiency halide perovskite solar cells

Nikolai Tsvetkov,[1,†] Byeong Cheul Moon,[1] Jeung Ku Kang,[1*]

Muhammad Ejaz Khan,[2,†] and Yong-Hoon Kim[2,*]

[1] Department of Materials Science and Engineering, Korea Advanced Institute of Science and Technology (KAIST), 291 Daehak-ro, Yuseong-gu, Daejeon 34141, Republic of Korea

[2] Department of Electrical Engineering Korea Advanced Institute of Science and Technology (KAIST), 291 Daehak-ro, Yuseong-gu, Daejeon 34141, Republic of Korea

* Corresponding author: jeungku@kaist.ac.kr and y.h.kim@kaist.ac.kr

† These authors contributed equally to this work.

**ABSTRACT:** Halide perovskite light absorbers have great advantages for photovoltaics such as efficient solar energy absorption, but charge accumulation and recombination at the interface with an electron transport layer (ETL) remains a major challenge in realizing their full potential. Here we report the experimental realization of a zipper-like interdigitated interface between a Pb-based halide perovskite light absorber and an oxide ETL by the PbO capping of the ETL surface, which produces an atomically thin two-dimensional metallic layer that can significantly enhance the perovskite/ETL charge extraction process. As the atomistic origin of the emergent two-dimensional interfacial metallicity, first-principles calculations performed on the representative $MAPbI_3/TiO_2$ interface identify the interfacial strain induced by the simultaneous formation of stretched I-substitutional Pb bonds (and thus Pb-I-Pb bonds bridging $MAPbI_3$ and $TiO_2$) and contracted substitutional Pb-O bonds. Direct and indirect experimental evidences for the presence of interfacial metallic states are provided, and a non-conventional defect-passivating nature of the strained interdigitated perovskite/ETL interface is emphasized. It is experimentally demonstrated that the PbO capping method is generally applicable to other ETL materials including ZnO and $SrTiO_3$, and that the zipper-like interdigitated metallic interface leads to about 2-fold increase in charge extraction rate. Finally, in terms of the photovoltaic efficiency, we observe a volcano-type behavior with the highest performance achieved at the monolayer-level PbO capping. The method established here might prove to be a general interface engineering approach to realize high-performance perovskite solar cells.

## 1. INTRODUCTION

With the maturation in the development of high-quality halide perovskite light absorbers, the control of their interfaces with charge extraction layers has emerged as a key to maximize the performance of perovskite solar cells (PSCs)[1–3]. Typical PSCs are fabricated by sequentially depositing an electron-transport layer (ETL), a light-absorbing perovskite layer with or without a mesoporous scaffold layer, a hole-transport layer (HTL), and a high work-function electrode onto a transparent conducting substrate. To improve the interfaces between perovskite absorbers and ETLs in PSCs, there have been great efforts such as the passivation of the perovskite layer by excess $PbI_2$[4] or fullerene interlayer[5] as well as the utilization of $SnO_2$ nanocrystals,[6] ZnO nanorods,[7] perovskite oxide $SrTiO_3$[8], graphene quantum dots/$SnO_2$,[9] graphene/$TiO_2$,[10] hierarchically structured $Zn_2SnO_4$,[11] acid-treated $TiO_2$,[12] black phosphorene-covered $TiO_2$[13] as ETLs. Still, realizing the perovskite/ETL interfaces with minimal defect concentrations remains an utmost challenge. For example, while the chlorine[14] or Pb oxysalt[15] passivation of a $TiO_2$/ETL interface has been proposed as optimal approaches for solution-processed PSCs, it is inadequate for mesoporous $TiO_2$ and other metal oxide transport layers that are most commonly used to realize high-efficiency PSCs. Hence, establishing a new interface structure that can meet this challenge is of enormous interest for the development of high-performance PSCs.

Meanwhile, pressure was recently identified as a powerful knob to fine-tune key properties of organic-inorganic hybrid halide perovskites[16,17] such as bandgap narrowing with carrier-life time prolongation,[18,19] multiple phase transitions with bandgap redshift and blue jump,[18,20,21] and even metallic state generation.[22–24] While the metallization was exclusively observed in ultrahigh-pressure bulk-phase cases, pressure or strain effects at the halide perovskite interfaces have been recently discussed[25–27] and were also shown to tailor the characteristics of low-dimensional halide perovskites[28,29].

In this work, we realize strain-induced two-dimensional (2D) metallic states at the zipper-like interdigitated atomically thin interfaces between oxide ETLs and Pb-based halide perovskites. Based on density functional theory (DFT) calculations, we predict that the controlled substitution of interfacial Ti atoms of $TiO_2$ ETL by Pb atoms at the $TiO_2$/methylammonium lead triiodide ($MAPbI_3$) interface or the establishment of Pb-I-Pb networks bridging $MAPbI_3$ and $TiO_2$ leads to metallic interfacial states. Experimentally, we realize the interdigitated methylammonium lead triiodide ($MAPbI_3$)/$TiO_2$ interface by capping $TiO_2$ with PbO and find that the monolayer-level PbO interdigitation significantly enhances the photovoltaic efficiency. The existence of metallic $Pb^0$ states at the Pb-capped $TiO_2$ ETL/$MAPbI_3$ interface, which could act as efficient mediators for electron transfer, is verified by XPS measurements. DFT calculations identify that the origin of a 2D metallic state at the interdigitated $MAPbI_3$/$TiO_2$ is the significant interfacial strain induced by the simultaneous formation of stretched I-substitutional Pb bonds and contracted Pb-O bonds. Indeed, based on transmission electron microscopy (TEM) measurements, we confirm the good agreement between the experimentally measured interfacial atomic geometries and DFT-predicted strained interfacial atomic structures. In addition, upon introduction of ant oxygen vacancy, we find that the structurally strained and electronically metallic nature of the interdigitated $MAPbI_3$/$TiO_2$ interface robustly suppresses oxygen vacancy effects, which contrasts the generation of deep trap states in the planar $MAPbI_3$/$TiO_2$ interface counterpart. Besides, we show that the synthesis method of capping ETL surfaces by a PbO monolayer is generally applicable to other oxide ETL materials such as ZnO and $SrTiO_3$ in combination with halide perovskite light absorbers including $MAPbI_3$ and triple-cation $Cs_{0.05}(FA_{0.85}MA_{0.15})_{0.95}Pb(I_{0.85}Br_{0.15})_3$ (CsFAMA). Furthermore, we reveal that electron injection and photovoltaic efficiency are maximized at the monolayer-level PbO capping of $TiO_2$ and subsequently decrease with the multilayer PbO capping, establishing a volcano-type correlation.

## 2. RESULTS AND DISCUSSION

Carrying out DFT calculations, we first establish the possibility of creating a 2D metallic layer by substituting the Ti atoms at the $TiO_2$ surface interfacing $MAPbI_3$ with Pb atoms or creating an interdigitated $TiO_2$/$MAPbI_3$ interface schematically shown in **Figure 1**a-b. By adopting a planar (101) anatase $TiO_2$ surface model, which matches the experimental situation discussed later, we prepared two additional surface models in which the Ti atoms in the top monolayer (100 %) and bilayer (200 %) are replaced by Pb atoms. Next, we completed the three interface models by placing a (110)-$MAPbI_3$ slab on the top of the planar, 100% Pb-substituted, and 200% Pb-substituted (101)-$TiO_2$ slab structures (**Figure 1**c left panel; see also Supporting Information Figures S1 and S2). The planer PbI-terminated perovskite surface has been employed, because $MA^+$- or $I^-$-terminated surfaces lead to a poor electronic connection between $TiO_2$ and perovskite layers[14,30,31].

To analyze the interfacial electronic structure, we first calculated the interfacial charge density differences according to

$$\Delta \rho = \rho_{MAPbI_3+(Pb-)TiO_2} - \rho_{MAPbI_3} - \rho_{(Pb-)TiO_2} \quad (1)$$

where $\rho_{MAPbI3+(Pb-)TiO2}$, $\rho_{MAPbI3}$, and $\rho_{TiO2}$ correspond to the charge densities of $MAPbI_3$+(Pb-substituted-)$TiO_2$ interface, $MAPbI_3$, and (Pb-substituted-)$TiO_2$ structures, respectively. We find that the interfacial charge transfer is significantly increased upon the Pb interdigitation (**Figure 1**c). This enhancement arises only up to the 100 % Pb substitution level, and the additional Pb introduction did not increase the interfacial charge transfer (Figure S2, Supporting Information).

In addition, as shown in **Figure 1**c right panel, the electronic properties of the interdigitated $MAPbI_3$/Pb-substituted $TiO_2$ heterostructures including the interfacial band alignment were analyzed using the layer-by-layer distributions of density of states (DOS).[32,33] Surprisingly, we find in the $MAPbI_3$/100 % Pb-substituted $TiO_2$ interface case the interdigitated interfacial layer becomes metallic (layer ⑤) and it additionally creates metal-induced gap states propagating into the neighboring $MAPbI_3$ with a thickness of about 4 Å (layers and ③ and ④). However, upon additionally substituting the Ti atoms in the next layer by Pb atoms (200 % Pb substitution case), we find that a small band gap opens within the second Pb-substituted $TiO_2$ or $PbO_2$ layer of a ~ 2.5 Å thickness (Figure S2, Supporting Information). The band mismatch between this PbO$_2$ layer and the net bulk TiO$_2$ has a type-II alignment, which can be expected to increase the interfacial resistance. Namely, DFT calculations predict a volcano-type correlation between the Pb substitution level and perovskite/ETL contact resistance with the optimum achieved at the perfectly interdigitated (100 % Pb substitution level) case.

To experimentally realize the interdigitated MAPbI$_3$/TiO$_2$ heterostructure, we have prepared the PbO-capped TiO$_2$ ETLs by treating mesoporous TiO$_2$ layers with a PbCl$_2$ aqueous solution.[34] By changing the solution concentration, we fabricated different ETLs with the Pb/(Ti+Pb) atomic ratios in the range between 3% and 40% as it was measured by XPS. We chose the particular ratios of 7%, 13%, and 24% (labeled as Pb7, Pb13, and Pb24 samples, respectively), corresponding to an incomplete PbO monolayer coverage, a complete PbO monolayer coverage, and a PbO multilayer coverage/nanocluster formation, respectively (Figure S3, Supporting Information). The XPS measurements show that Cl anions are absent on the TiO$_2$ surface (Figure S4, Supporting Information), indicating that the changes in the electronic structure of the perovskite/TiO$_2$ interface will be due to the deposited Pb cations. Next, to check the possibility of dipole formation at the TiO$_2$ surface with the PbO treatment, we performed ultra-violet photoelectron spectroscopy measurements (Figure S5, Supporting Information) to evaluated the effect of PbO capping on the TiO$_2$ work function value. If change of the work function is taking place this could be a fingerprint of dipole formation.[35] However, we did not observe any significant change in secondary offset position. Thus, we conclude that PbO capping-induced dipole formation does not take a place.

On top of PbO-capped TiO$_2$ surfaces with different PbO capping ratios, we then deposited MAPbI$_3$ layers and evaluated the efficiency of interfacial charge injection from the MAPbI$_3$ perovskite to the PbO-capped TiO$_2$ ETL through the time-resolved photoluminescence (TRPL) measurement. We present in **Figure 1**d the TRPL decay curves of PL peaks at 775 nm upon laser excitation for the control TiO$_2$/MAPbI$_3$ and MAPbI$_3$/monolayer PbO/TiO$_2$ (Pb13). Those for the MAPbI$_3$/incomplete-monolayer PbO/TiO$_2$ (Pb7) and MAPbI$_3$/multilayer PbO/TiO$_2$ (Pb24) are also shown in Figure S6 (Supporting Information). Using the triple exponential functions to extract carrier lifetime[36], the TRPL decay curves indicate the much faster TRPL signal decay for the Pb13 sample compared to that for the reference non-PbO-capped TiO$_2$/MAPbI$_3$. Also, the fastest component that reflects the electron injection rate from MAPbI$_3$ to the ETL decreases from 8 ns to 4 ns as TiO$_2$ is capped with a PbO monolayer. We observed negligible differences in morphology and crystal structure of perovskite layers deposited on non-PbO-treated and Pb13 PbO-capped TiO$_2$ from X-ray diffraction (XRD) and scanning electron microscopy (SEM) (Figure S7, Supporting Information), which allows us to attribute the observed differences to the change in the electronic structure of the perovskite-TiO$_2$ interface induced by the PbO capping of TiO$_2$. Accordingly, we conclude that the electron extraction from a perovskite light absorber becomes more efficient in the presence of a Pb interdigitation through a metallic interfacial layer that plays the role of an efficient charge transfer mediator (**Figure 1**e).

Having confirmed the theoretically predicted enhancement of charge injection efficiency at the interdigitated perovskite/ETL interface, we analyzed the PbO-capped surfaces and interfaces in more detail. The high-angle annular dark field-scanning TEM (HAADF-STEM) images of a TiO$_2$ nanoparticle treated with the PbO capping are shown in **Figure 2**a-b. The brighter dots in the images indicate the strong diffraction from the heavy Pb atoms, while the less bright dots stand for the weak diffractions from the lighter Ti atoms. We observe a uniform bright signal from the surface of TiO$_2$ nanoparticles, indicating the almost complete surface coverage with Pb atoms. The thickness of the PbO layer was determined to be of about 0.4 nm. Within the nanoparticle bulk we observe the typical TiO$_2$ lattice spacing of 0.35 nm corresponding to the preferred (101) orientation of the anatase TiO$_2$ surface. The high-resolution STEM analysis (Figure S3, Supporting Information) shows that Pb atoms are located at the TiO$_2$ lattice fringes. This indicates that the Pb atoms are incorporated into the TiO$_2$ lattice at the surface. The additional increase of Pb loading led to the appearance of PbO clusters at the TiO$_2$ surface, where the inter-planar distance of about 0.30 nm between lattice fringes matches with that of the (110) PbO clusters.

The XPS measurement data of the Pb7, Pb13, and Pb24 TiO$_2$ samples are shown in Figure S8, Supporting Information. They provide the information on the oxidation state of the Pb atom at the TiO$_2$ surface. The inorganic B-site cations such as Pb or Sn within halide perovskites are in the bivalent oxidation state (2+), and the higher oxidation-state sites at the interface can act as recombination centers. The main Pb 4f$_{7/2}$ peaks from all the samples are located at around 139 eV, confirming that a Pb oxidation state is 2+. However, for the P24 sample, the deconvolution of Pb 4f$_{7/2}$ spectrum (Figure S9, Supporting Information) shows an extra shoulder in a lower binding energy region, indicating the presence of $Pb^{3+}$. The partial trivalent oxidation state of Pb cations can be understood in terms of the above-mentioned appearance of PbO clusters at the $TiO_2$ surface. Similarly, the Fourier-transform infrared (FTIR) spectra (Figure S10, Supporting Information) reveal a peak at around 1350 cm$^{-1}$ corresponding to the Ti-O-Pb bond vibration in all the Pb7, Pb13, and Pb24 structures.[37,38] However, a shoulder peak at around 1000 cm$^{-1}$ matching with the Pb-O-Pb bond vibration of lead oxide[38] was observed only in the Pb24 sample with nanoclusters.

Moving on to the analysis of perovskite/PbO-capped $TiO_2$ heterojunctions, we show in **Figure 2**c-d the energy-dispersive X-ray spectroscopy scanning transmission electron microscopy (EDS-STEM) data together with the STEM images of the interfaces given in insets. **Figure 2**c shows that I atoms in a non-PbO-capped $TiO_2$/MAPbI$_3$ interface penetrates deeper than Pb atoms, indicating that the bonding between $TiO_2$ and MAPbI$_3$ is taking place through the formation of Ti-I bonds, which is in line with previous reports[31]. Meanwhile, the EDS measurement of the Pb13 sample with the monolayer-level PbO capping (**Figure 2**d) reveals that Pb atoms penetrate deeper into the $TiO_2$ side of the interface compared to I atoms. Moreover, the difference in the penetration depth between those two elements is of around 0.5 nm, corresponding well to the theoretically estimated thickness of a perfectly interdigitated $TiO_2$/MAPbI$_3$ interface (details will be discussed below). Thus, the TEM, XPS, and FTIR analyses verify that the prepared Pb7-, Pb13-, and Pb24-based $TiO_2$/MAPbI$_3$ heterojunctions lead to the incomplete monolayer, near-complete monolayer, and multilayer or cluster-form PbO coverages of the $TiO_2$ surface.

Because the theoretically identified interfacial metallicity at the interdigitated $TiO_2$/MAPbI$_3$ interface was accompanied by significant rearrangements of atomic structures (**Figure 1**c), we scrutinized whether the atomic structures at the hetero-interfaces obtained from our experiment (**Figure 3**a-b) match those from computation (**Figure 3**c). First, the DFT-optimized structure of the monolayer-level Pb interdigitated $TiO_2$(101)/MAPbI$_3$(110) model (**Figure 3**c) exhibits the inter-layer thickness distribution of 0.44 nm (cyan left arrow) to 0.22 nm (white left arrow) and to 0.66 nm (orange left arrow), which is noticeably different from that of its planar counterpart, 0.35 nm-0.29 nm-0.64 nm (Figure S11, Supporting Information). Experimentally, we performed high-resolution TEM measurements on the Pb13 sample-based $TiO_2$/MAPbI$_3$ interface and determined the change of the inter-planar distances across the interface by monitoring the evolution of the TEM signal along the interface-normal direction as depicted by white arrows (**Figure 3**a). The peaks in the TEM signal intensity spectrum refer to the positions of cations in the lattice and the distance between the peaks reflects the inter-planar distance between atomic layers. Specifically, the TEM signals from bulk $TiO_2$ and MAPbI$_3$ regions show the about 0.35 nm interlayer spacing of (101) $TiO_2$ as well as the around 0.60-0.65 nm interlayer spacing of (100) MAPbI$_3$. Meanwhile, in the zoomed-in **Figure 3**b, the inter-planar distance profile along the interface-normal direction shows a transition from 0.45 nm (cyan left arrow) to 0.22 nm (black left arrow) and to 0.65 nm (orange left arrow), which is in an excellent agreement with the computationally obtained values (0.44 nm-0.22 nm-0.66 nm) (**Figure 3**c).

Because apparently the Pb interdigitation-induced interfacial strain is the origin of 2D metallicity, we additionally examined the interfacial-region bond length changes in DFT-optimized zipper-like interdigitated $TiO_2$/MAPbI$_3$ interface structures. As summarized in **Figure 3**d, we find that the replacement of inter-layer Ti-I bonds by Pb-I bonds results in ~ 0.4 Å elongation (from 2.92 Å to 3.32 Å of the average bond length). This is then accompanied by the ~ 0.34 Å compression of neighboring inter-layer O-Pb bonds (from 2.62 Å to 2.28 Å of the average bond length). These quantify the level of significant strain and resulting structural distortions at the interdigitated $TiO_2$/MAPbI$_3$ interface.

Having identified the structural deformation accompanied by Pb-I-Pb bonds extending from MAPbI$_3$ into $TiO_2$ as the origin of the 2D interfacial metallicity, we additionally explored a more direct evidence of the presence of metallic states in the prepared PbO-capped *mp*-$TiO_2$ layer/MAPbI$_3$ perovskite heterostructures through the XPS analysis. The XPS Pb 4f core-level spectra recorded from $TiO_2$/MAPbI$_3$ interfaces without and with PbO capping are shown in **Figure 3**e, and the survey spectra with similar Ti and Pb signal intensities for different samples are displayed in Figure S12 (Supporting Information). For the $TiO_2$/MAPbI$_3$ hetero-interface without capping, only the peaks corresponding to $Pb^{2+}$ were detected. However, for the PbO-capped $TiO_2$/MAPbI$_3$ counterparts, additional peaks shifted by about 1.7eV from the main peaks to the lower binding energy region are observed. These features correspond to the metallic state of Pb,[39] which is another evidence of the 2D metallic states at the PbO-capped $TiO_2$/MAPbI$_3$ interfaces.

To provide additional experimental evidence of the emergence of metallic states at the interdigitated

TiO$_2$/MAPbI$_3$ interface, we fabricated the non-PbO-capped and PbO-capped TiO$_2$/perovskite heterojunctions and performed in-plane current-voltage (I-V) measurements in the dark condition (**Figure 4**a left top panel). Without including bulk perovskite and HTL layers in the full PSC configuration, it is expected that the influence of interfacial metallic states will be amplified in these measurements. We indeed observe that, while the non-PbO-capped TiO$_2$/perovskite interface exhibits a typical semiconducting I-V behavior, the PbO-capped counterpart shows an Ohmic-like I-V curve or a signature of metallic states (Figure S13, Supporting Information).

Next, using the same device sample, we performed the in-plane electrochemical impedance spectroscopy (EIS) analysis and explored the correlation between the Pb-interdigitation-induced metallicity and the interfacial defect density.[40,41] The EIS spectra were fitted using the circuit shown in the left bottom panel of **Figure 4**a consisting of a series resistance and a capacitive constant-phase element (CPE) connected in parallel. In this model, the chemical capacitance of the system can be extracted according to[42]

$$C_{Chem} = Q \cdot w_{max}^{1-n} \quad (2)$$

where Q and n are parameters of CPE element in equivalent circuit (**Figure 4**e) and $w_{max}$ is the angular frequency of the highest point of the semicircle. We estimated that the chemical capacitance of the configuration with the non-PbO-capped interface is around 25% higher compared to the Pb13 PbO-capped TiO$_2$ counterpart. It is known that the defect density and its related interface capacitance at TiO$_2$/perovskite interface are closely related to the creation and migration of oxygen vacancies.[41,43,44] Thus, our experimental results of decreased interface capacitance clarify the passivation of oxygen vacancy and other interfacial defects by the PbO capping of TiO$_2$ or the formation of interdigitated TiO$_2$/perovskite interfacial structures (**Figure 3**).

To provide the microscopic understanding of experimental results, we then theoretically studied oxygen vacancy V$_O$ defects formed at both planar and interdigitated TiO$_2$/MAPbI$_3$ interfaces.[14,15,43,44] Considering different V$_O$ sites in the interfacial region (Figure S14, Supporting Information), we found from **Figure 4**b that both deep and shallow defect states are generated in the planar interface case. Particularly deep defect levels are originated from the dangling bonds of Ti atoms at the V$_O$ site (Figure S15, Supporting Information). In contrast, for the corresponding Pb-interdigitated interface cases, **Figure 4**c shows that the V$_O$-originated defect states are suppressed. This behavior is different from the typical defect passivation methods reported in the literature because in our case extrinsic defect-passivating materials have not been introduced or the zipper-like interdigitated TiO$_2$/MAPbI$_3$ interface is still a "pristine" interface. Instead, the layer-by-layer projected DOS (**Figure 4**d) and the local DOS of V$_O$ states (**Figure 4**e) show that the oxygen defect states are washed out by metallic 2D interfacial states. Apparently, the effective suppression of defect states in the interdigitated TiO$_2$/MAPbI$_3$ interface is originated from the defect-tolerant nature of already much distorted interfacial atomic structures, which contrasts the situation in the planar TiO$_2$/MAPbI$_3$ interface case where the appearance of deep V$_O$ levels was accompanied by a noticeable level of structural distortions.

Finally, to evaluate the influence of strain-engineered metallic interfacial layer on the energy conversion efficiency in complete devices, we fabricated PSCs based on the mesoporous TiO$_2$ interfaces with different PbO capping levels and determined their photovoltaic performance under the simulated AM 1.5 G illumination at 100 mW/cm$^2$. The photovoltaic parameters of the devices have been reproduced on 15–20 cells for each type of PSCs and we confirmed that there is no significant difference observed in the J-V curves with different scan rates. In addition, we have found the marginal hysteresis in PSCs with Pb-capped ETLs. The deviated efficiency between forward and reverse scans (Figure S16, Supporting Information) is only about 1-1.5%. The current-voltage (J-V) curves of the best performing devices fabricated with CsFAMA perovskite light absorber are shown in Figure S17 (Supporting Information) and the corresponding photovoltaic parameters of PSCs are summarized in Table S1 (Supporting Information). The PSCs based on ETLs with monolayer capping show superior performance compared to those based on reference ETL.

We attribute the observed increase of FF and V$_{oc}$ of the PSCs with Pb-capped ETLs to the interfacial metallic states-induced reduction of carrier recombination at the interdigitated TiO$_2$/perovskite interface (Table S1, Supporting Information). As discussed earlier, the defect suppression at the interdigitated TiO$_2$/perovskite interface is an important consequence of its structurally distorted and electronically metalized nature. To quantify this feature, additional measurements of the current-voltage behavior in dark conditions (Figure S18, Supporting Information) were performed. Indeed, we observe the decrease of the leakage current in the dark for PSCs with PbO capped ETL, which is the evidence of reduced defect concentration or defect suppression.

We performed the EIS measurements for the PSCs with different ETLs and show in **Figure 5**a in the Nyquist plots together with the equivalent circuit used for fitting. The two semicircles are observed in the EIS spectra, and while the first semicircle in the high-frequency range is attributed to the charge transfer resistance at the ETL-perovskite and perovskite-HTL interfaces[45,46] and the second semicircle in the low-frequency range is ascribed to the charge transfer within the $TiO_2$ layer. In the equivalent circuit, $R_s$ corresponds to the series resistance of the cell, which is determined from the high-frequency intercept of the left semicircle with the x-axis, and $R_{TiO2}$ and $CPE_{TiO2}$ refer to the charge transfer resistance and the CPE of the $TiO_2$ layer, respectively. $R_{ct}$ and $CPE_{ct}$ represent the charge transfer resistance and the capacitance at the ETL-perovskite and perovskite-HTL interfaces, respectively. We find that the $R_s$ values are similar for PSCs with and without PbO-capped ETLs, indicating that the difference in the device performance is primarily affected by the charge transfer resistance at the interfaces. We have observed a good correlation between $R_{ct}$ and $V_{oc}$ values of PSCs (**Figure 5**a-b). As PSCs have been fabricated with the identical perovskite/Spiro-MeOTAD interfaces, the difference in the EIS spectra is attributed to the change in $R_{ct}$ at the $TiO_2$/perovskite interface. We also observe that $R_{ct}$ is decreasing with increasing the PbO coverage up to the monolayer level, but afterwards it is significantly increasing, confirming that the PbO coverage on the $TiO_2$ surface only up to the monolayer limit is favorable for the ETL/perovskite interfacial charge transfer. Utilizing the $MAPbI_3$ and the triple-cation CsFAMA perovskite light absorbers, we obtained similar performance improvement trends that show the volcano-type behaviors as shown in **Figure 4**c (Tables S1 and S2, Supporting Information). Moreover, we find that the PSC efficiency increases from 18.7% with a reference $TiO_2$ ETL up to 21.2% with the monolayer PbO capping-level Pb13 ETL. The further increase of the PbO capping density is shown to result in the dramatically decreased PSC efficiency down to 16.7%.

Additionally, we explored the generality of the Pb-capping approach by adopting $SrTiO_3$ and ZnO ETLs. Figure S19 (Supporting Information) verifies that $V_{oc}$ and FF are enhanced with the Pb monolayer-capped ETL layers. However, we did not observe any significant improvement in PSC performance with Sn-capped ETLs (Figure S20, Supporting Information). In a theoretical calculation that will be detailed elsewhere, we considered a Pb-interdigitated $MAPbI_3/SnO_2$ interface but could not find metallic interfacial states. This again emphasizes that the formation of the Pb-I-Pb bonds extending from $MAPbI_3$ into $TiO_2$ or the creation of a zipper-like interdigitated $TiO_2/MAPbI_3$ interface is the source of the emergent interfacial 2D metallicity.

Finally, we have also investigated the stability of PSCs fabricated with bare and Pb-capped $TiO_2$ ETLs. **Figure 4**d reveals that after 13 days the PSC with Pb-capped ETLs preserves more than 95% of the initial power conversion efficiency. In comparison, with the PSCs fabricated with bare ETLs, performance declined to an efficiency less than 80% of the initial performance. While the improved stability of the devices with Pb-capped ETLs can be also related to the process of $PbCl_4$ treatment that can lead to the removal or suppression of $TiO_2$ surface defects such as oxygen vacancies,[47,48] we suggest that even at the same level of defect density the interdigitated perovskite/ETL interface will be more stable than planar interface due to the effective passivation of defect states by the strain-induced 2D metallic states extensively discussed above (**Figure 4**).

## 3. CONCLUSION

In summary, we demonstrated a strategy to implement a zipper-like interdigitated ETL/halide perovskite interface by controllably capping the surface of an oxide ETL with PbO and subsequently growing a Pb-based halide perovskite light absorber. We showed that the method achieves an efficient electron extraction and a high photovoltaic efficiency, and they result from the metallization of the interdigitated ETL/perovskite interface. As the atomistic origins of the 2D metallicity emerging at the interdigitated $TiO_2/MAPbI_3$ interface, DFT calculations identified the strain induced by the replacement of original Ti-I bonds by much longer Pb-I bonds. The formation of Pb-I-Pb bonds bridging $TiO_2$ and $MAPbI_3$ in turn notably shortens the nearby O-Pb bonds, and the resulting strain significantly distorted the interfacial atomic structure. The existence of a metallic layer at the interdigitated $TiO_2(101)/MAPbI_3(110)$ interface was directly and indirectly confirmed by the XPS and high-resolution cross-sectional TEM measurements, respectively. The non-conventional defect-passivating nature of the strain-induced 2D interfacial metallic layer was revealed. In addition, the TRPL measurements of the perovskite/PbO-capped ETL interfaces showed an about 2-fold increased electron extraction rate compared to that of the bare perovskite/ETL interface. Accordingly, the enhancements in open circuit voltage and fill factor were achieved, but the improvements were seen only up to the monolayer PbO capping level. The additional excessive PbO capping resulted in an increased interfacial resistance and a decreased PSC efficiency, establishing a volcano-type trend. The creation of a 2D electron liquid at halide perovskite-based interfaces might prove to be a general strategy to improve interfaces for high-performance hybrid halide perovskite solar cells.

**EXPERIMENTAL SECTION**

**Computational details and model structures.** First-principles DFT calculations were performed within the project-augmented wave scheme[50] and generalized gradient approximation in the Perdew-Burke-Erzenhof form revised for solids (PBEsol)[51] as implemented in the Vienna Ab initio Simulation Package (VASP).[52] The plane-waves were expanded with a kinetic energy cutoff of 400 eV to obtain basis sets. Atomic structures were optimized using the conjugate-gradient approach until the Hellmann–Feynman forces were less than 0.025 eV/Å. Weak dispersive interactions were taken into account with the DFT-D2 method.[53] Starting from the anatase $TiO_2$ and $MAPbI_3$ models obtained previously,[54-56] we constructed the planar $TiO_2(101)/MAPbI_3(110)$ interface model by depositing a 3×5 pseudocubic $MAPbI_3(110)$ on a 5×3 anatase $TiO_2(101)$ slab with lattice parameters $a$ = 18.92 Å and $b$ = 30.72 Å.[57] Along the direction normal to the interface, together with two unit-cell-thickness $MAPbI_3(110)$ slabs, we included two (three) octahedral layers of $TiO_2(101)$ to construct planar and monolayer (bilayer) Pb-substituted heterojunction models. Considering the experimental situation of growing $MAPbI_3$ on top of $TiO_2$ and the softer nature of $MAPbI_3$, we fixed the supercell lattice parameters to the experimental value of $TiO_2$. Next, a vacuum space of 15 Å was included to avoid artificial interactions with neighboring supercell images within the periodic boundary conditions. More computational details can be found in our previous reports.[54-56]

**Fabrication of perovskite solar cell devices.** The FTO-etched glasses have been cleaned sequentially with DI water, ethanol and acetone for 15 minutes by using the ultrasonic bath. Substrates were treated using the UV-ozone plasma during 20 minutes. The $TiO_2$ blocking layer was deposited via spin-coating at velocity of 2000 rpm for 60 second using 0.3 M titanium diisopropoxide bis(acetylacetonate) in 1-butanol solution that was heated afterward at 125 °C for 5 minutes. Then, the substrates were annealed at 450 °C for 30 minutes. We also used spray pyrolysis as an alternative method to deposit the compact $TiO_2$ blocking layer. Then, the $TiO_2$ paste in ethanol (1:9 weight ratio) solution was spin-coated onto the substrate at 4000 rpm for 60 seconds and annealed at 125 °C for 5 minutes. Subsequently, the substrates were annealed at 450 °C for 30 minutes followed by the treatment in 40 mM $TiCl_4$ aqueous solution at 70 °C for 30 minutes. Next, substrates were washed using DI water and ethanol. Then, there were blow-dried with argon and annealed at 450 °C for 30 minutes. To deposit PbO, $TiO_2$ film layers were dipped into the aqueous $PbCl_2$ solution at 90 °C for 60 seconds and washed with water and ethanol. Next, the substrates were annealed at 450 °C for 30 minutes in air. The Pb contents on the surface of $TiO_2$ was controlled by changing the concentration of $PbCl_2$ solution varied from 2 mg/L to 12 mg/L. Also, the perovskite deposition was conducted inside the nitrogen filled glovebox with a moisture level of 10-15 ppm. For devices with highest efficiency, we used the triple cation perovskite with a nominal composition of $Cs_{0.05}(FA_{0.85}MA_{0.15})_{0.95}Pb(I_{0.85}Br_{0.15})_3$[49]. The 1.5 M perovskite solution was prepared by dissolving in DMF:DMSO mixture (4:1 v/v) with stoichiometric amount of FAI, MABr, $PbI_2$ and $PbBr_2$. All powder reagents were purchased from TCI chemicals (Japan) and used without further purification. Finally, the appropriate amount of 1.5M solution of CsI dissolved in DMSO was added to the perovskite solution. After overnight stirring, the solution was deposited on the $TiO_2$ layer by spin coating at 500 rpm for 10 sec and at 6000 rpm for 25 sec sequentially. Next, the 250 μl of chlorobenzene solution was dropped at the film surface. Then, the substrates were annealed at 100 °C for 25 minutes. As regards to a hole transporting material, the Spiro-MeOTAD solution was deposited by spin-coating at 4000 rpm for 30 seconds. The solution was prepared by dissolving 72.3 mg of Spiro-MeOTAD powder (Lumtec, Taiwan) into the mixed solution of 28.8 μl 4-tert-butyl pyridine, 17.7 μl stock solution with 52 mg of Li-TFSI in 100 μl acetonitrile, and 1 ml of chlorobenzene. Finally, the 50 nm-thick gold layer was deposited using the thermal evaporator for the back electrode of a photovoltaic solar cell with a deposition rate of 3-4 Å/s.

**Device characterizations.** The TEM measurements of structures were performed using the double Cs-corrected transmission electron microscope Titan3 G2 60-300, Jeol. The photovoltaic parameters of solar cell devices were measured by the solar simulator (Newport, Oriel) with the potentiometer (Compactstat, IVIUM). The devices were measured under ambient air using the aperture mask with the active area of 0.105 $cm^2$. The I-V curves were recorded with the scan rate of 0.02 mV/s. The EIS measurements were performed under open circuit condition and 100 mW/$cm^2$ illumination in the frequency range from 1 MHz to 1 Hz with the IVIUM potentiometer, and the amplitude of the modulated voltage was 10 mV. The CBM positions were analyzed with the XPS instrument (K-Alpha, Thermo Scientific). Calibration of the spectra was performed by referring to the C 1s peak (C-C bond) at 285.0 eV. The VBM positions were estimated by extrapolation of the leading edge in valence band spectra. The XPS measurements were performed with the 90° take-off angle giving the ~6 nm probing depth (depth from where 90% of all signal is coming, 3x IMPF) in TiO$_2$. To perform XPS measurements at the TiO$_2$/perovskite interface the capping perovskite layer from the cells using razor blade thus the mp-TiO$_2$/perovskite interface was exposed to X-ray beam. For different samples the determined Pb/Ti ratios were very similar indicating that the ratio in the interface signal in the whole photoelectron yield is similar for different samples investigated. The REELS spectra were obtained using the Auger electron spectroscopy (AES; VG micro-lab 350 system). Furthermore, the TRPL measurements were conducted using the time-correlated single photon counting (TCSPC) method using a fluorescence lifetime spectrometer (FL920, Edinburgh Instruments). The tri-exponential function was used for decay-fitting of the PL decay curve based on the following equation of

$$f(t) = \sum_i A_i \exp(-t/\tau_i) + B \qquad (3)$$

where $A_i$ is the decay amplitude, $\tau_i$ is the decay lifetime, B is the constant and i is the exponential constant value used for decay fitting of the PL decay.

## ACKNOWLEDGMENTS

This research was supported mainly by the Global Frontier R&D Center for Hybrid Interface Materials (2013M3A6B1078884) and the National Research Foundation of Korea (2017R1A2B3009872, 202019M3E6A1104196, 2020R1A4A2002806, 2020H1D3A2A02104083). Computing resources were provided by the KISTI Supercomputing Center (KSC-2018-CHA-0032).

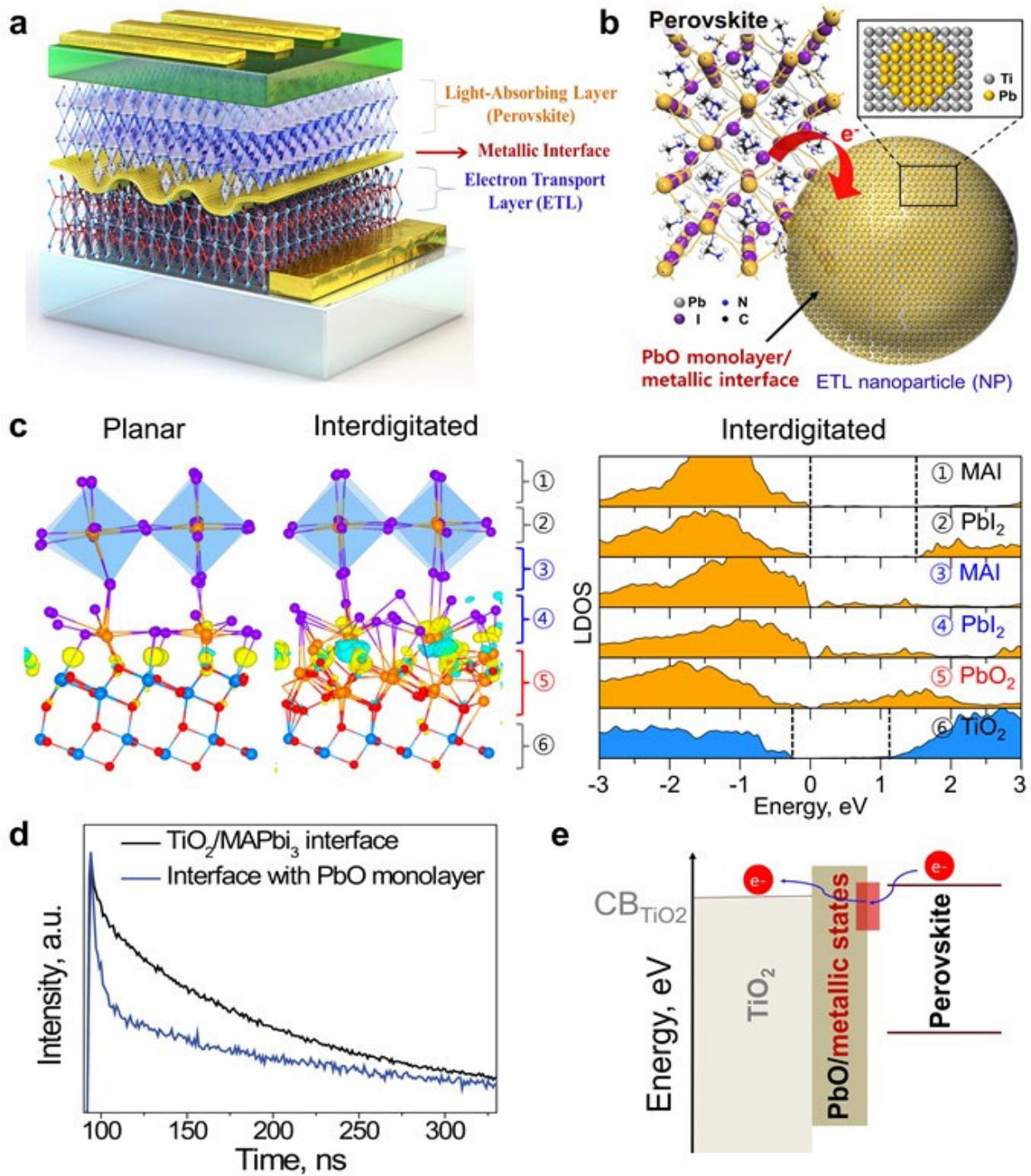

**Figure 1. Metallic states and charge transfer at perovskite/TiO$_2$ interface.** (a) Schematics of a PSC with a PbO-capped mp-TiO$_2$ ETL. (b) Illustration of electron injection process from a perovskite to a PbO-capped TiO$_2$ ETL. (c) Left panel: DFT-derived atomic structures of the planar and interdigitated (110)-TiO$_2$/(110)-MAPbI$_3$ interfaces. Overlaid are the interface-originated charge density differences in each case. Isosurface values are +0.005 eV/Å$^3$ and −0.005 eV/Å$^3$ for the charge accumulation (yellow) and depletion (cyan), respectively. Right panel: Layer-by-layer projected DOS for the Pb-interdigitated TiO$_2$/MAPbI$_3$ model. The left and right dotted lines indicate the valence and conduction band edges of semiconducting regions, respectively. (d) TRPL decay curves of the MAPbI$_3$ PL peaks obtained from TiO$_2$/MAPbI$_3$ interface without and with PbO capping. (e) Proposed mechanism of the enhancement of charge transfer across the interface between perovskite and a PbO-capped TiO$_2$ layer having metallic states.

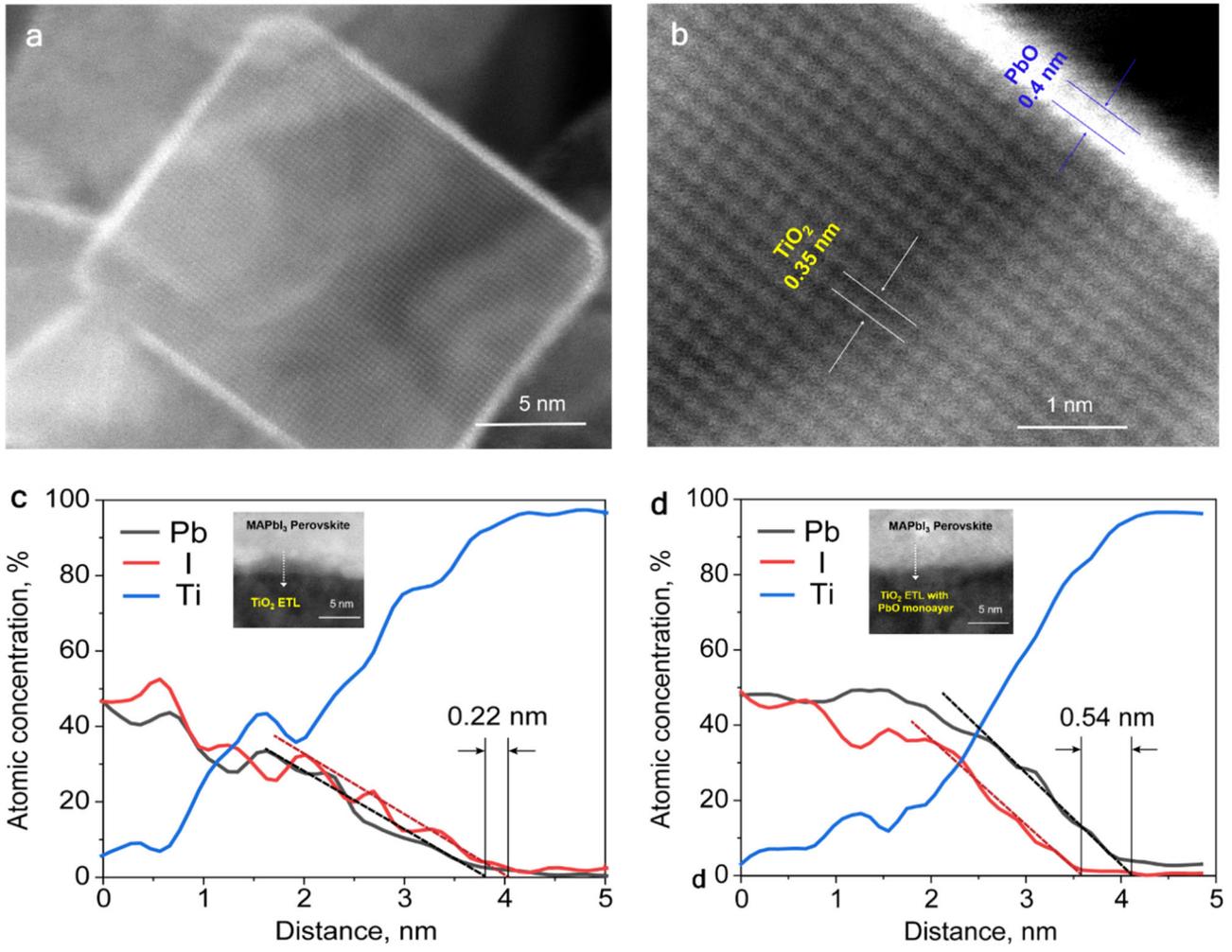

**Figure 2. Characterization of PbO monolayer at TiO$_2$ surface and at interface with MAPbI$_3$.** (a) High-resolution STEM image of the TiO$_2$ nanoparticles capped with a PbO monolayer. (b) Magnified image of the lattices near a particle edge, where the more intense color corresponds to the stronger diffraction from heavy Pb atoms compared to lighter Ti atoms. EDS elemental line profiles for Ti, I and Pb elements of (c) TiO$_2$/perovskite and (d) TiO$_2$/PbO monolayer/perovskite interfaces along the dotted arrows as shown in inserts.

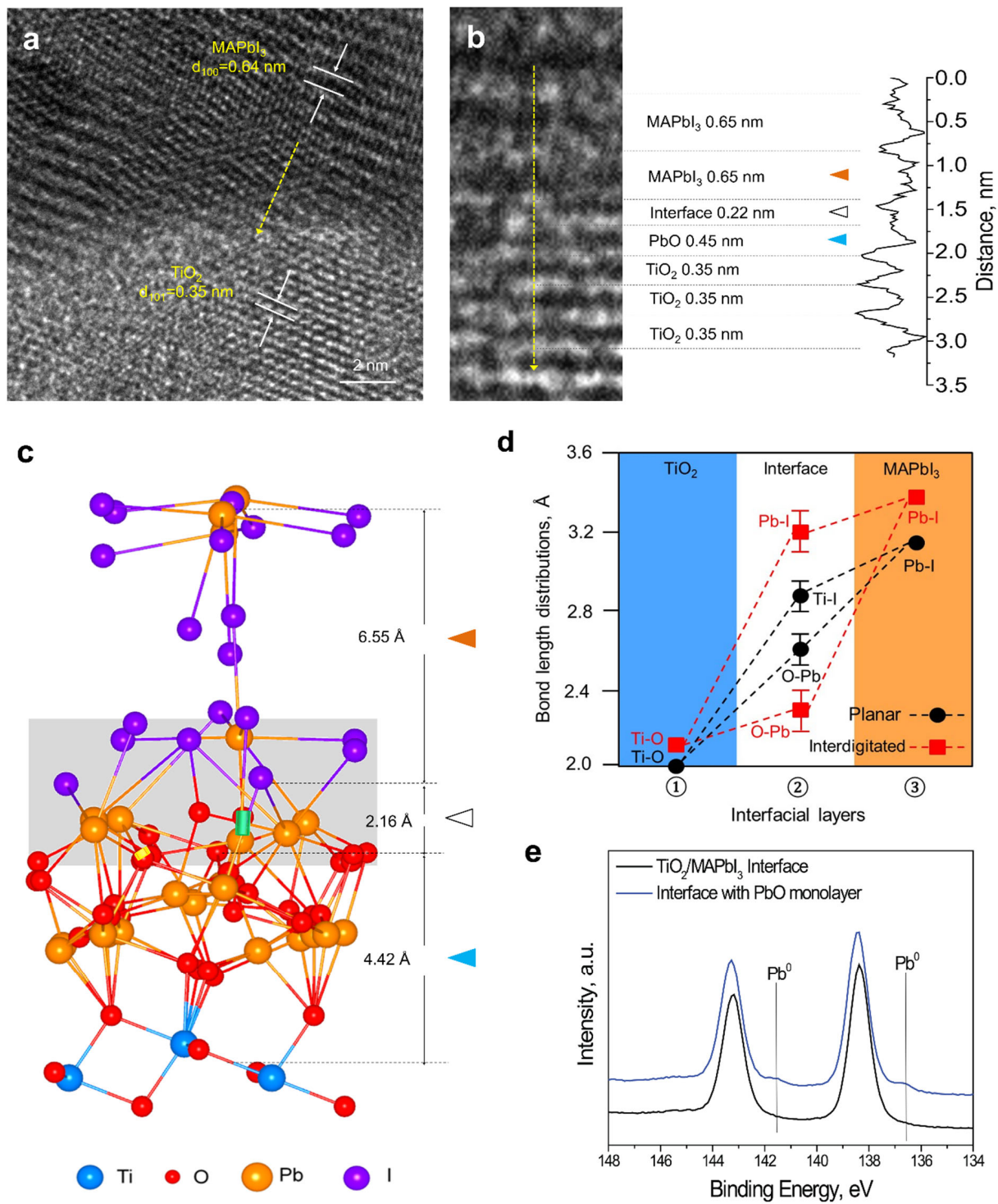

**Figure 3. Evidence and mechanisms of the strain-induced metallic states at TiO₂/PbO/MAPbI₃ interface.** (a) High-resolution TEM image of a TiO₂/PbO monolayer/perovskite interface. (b) Magnified TEM image and signal intensity change along the white arrow in (a). (c) DFT optimized atomic structure of a TiO₂(101)/PbO/MAPbI₃(110) interface, where green and yellow cylinders indicate the representative interfacial Pb-I and O-Pb bonds, respectively. The grey shadow box indicates the region where Pb-I and O-Pb bond lengths in (d) were measured. (d) Bond length distributions across the interdigitated TiO₂(101)/PbO/MAPbI₃(110) interface, where the numbers 1, 2 and 3 correspond to the layers marked with orange, black and blue left triangles in (b) and (c). (e) XPS spectra for the Pb 4f peak recorded for the interfaces between MAPbI₃ and mp-TiO₂ with and without PbO monolayer capping, where the small Pb⁰ peaks confirm the presence of metallic states at the TiO₂/PbO/MAPbI₃ interface.

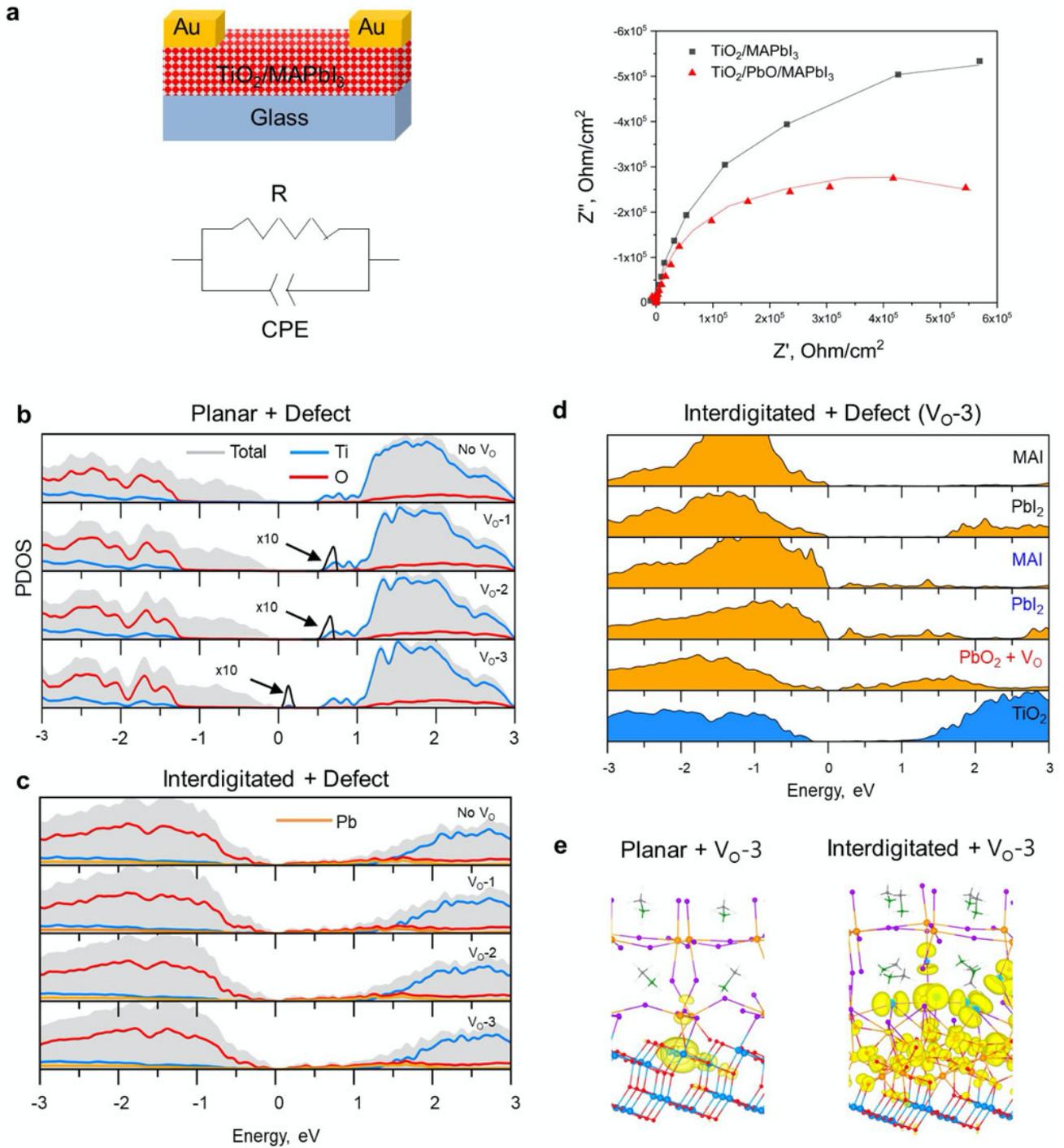

**Figure 4. Effects of defects at the interdigitated TiO$_2$/perovskite interface.** (a) The EIS data (right) recorded from the in-plane configuration (left top) of the mesoporous TiO$_2$/perovskite heterojunctions prepared with and without PbO capping. An equivalent circuit used for fitting is also shown (left bottom). DFT-calculated projected DOS for the (b) planar and (c) interdigitated TiO$_2$/MAPbI$_3$ interface models with an O vacancy introduced at three different interfacial sites. For the planar interface case, depending on the O vacancy position, we obtain shallow (V$_O$-1 and V$_O$-2) and deep defect levels (V$_O$-3). Note the disappearance of localized defect states in interdigitated TiO$_2$/MAPbI$_3$ interface models. (d) Layer-by-layer projected DOS of the interdigitated TiO$_2$/MAPbI$_3$ model with an oxygen vacancy that originally produced deep defect levels (V$_O$-3). Interfacial metallic states effectively wash out localized O vacancy defect states. (e) The local DOS plots of the V$_O$-3 defect states that appear as deep defect states in the planar TiO$_2$/MAPbI$_3$ interface (left panel) and buried into the delocalized metallic states in the interdigitated TiO$_2$/MAPbI$_3$ counterpart (right panel). Isosurface values are 0.002 eV/Å$^3$ and 0.002 eV/Å$^3$ for the charge accumulation (yellow) and depletion (cyan), respectively.

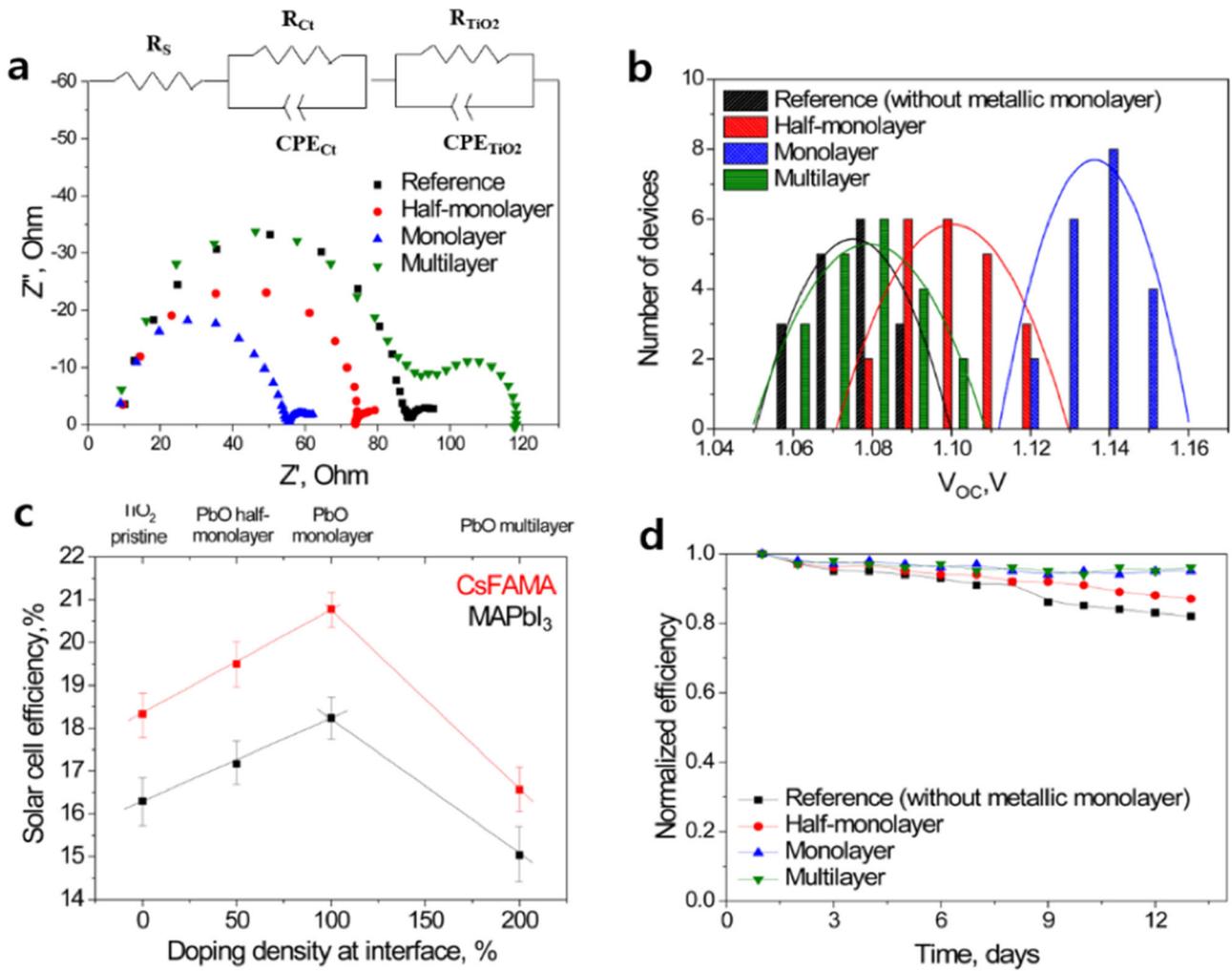

**Figure 5. Performance of PSCs with different PbO-capped ETLs.** (a) Electrochemical impedance spectra of PSCs with bare $TiO_2$ and various PbO-capped ETLs obtained under 1.5 AM illumination and open circuit conditions. (b) $V_{oc}$ parameters of PSCs fabricated with different ETLs. (c) Summary on the solar cell efficiencies of PSCs with $MAPbI_3$ and CsFAMA perovskites with different Pb doping densities on different doping densities at the interfaces. (d) Normalized efficiencies versus days for PSCs having different PbO cappings and kept under ambient conditions with 20%–30% humidity at room temperature.

**SYNOPSIS TOC**

The formation of zipper-like interdigitated perovskite/ETL interface by the PbO capping of $TiO_2$ results in interfacial metallic states owing to the strain induced by the formation of stretched I-substitutional Pb bonds and contracted substitutional Pb-O bonds. The structurally strained and electronically metallic nature of the interdigitated interface robustly suppresses defect states and leads to about 2-fold increase in charge extraction rate and a volcano-type solar cell efficiency behavior with the maximum at the perfectly interdigitated configuration.

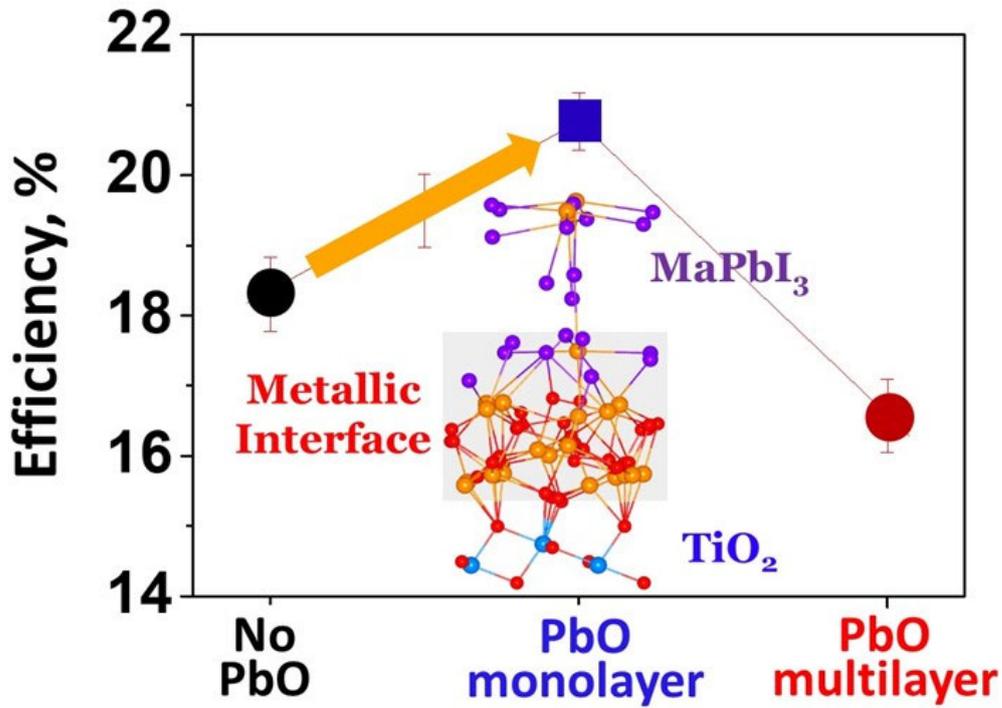